\documentclass[12pt,preprint]{aastex}
\usepackage{emulateapj5}
\usepackage{onecolfloat}
\usepackage{apjfonts}
\usepackage{subfigure}	

\newcommand{\inve}{\'e}

\newcommand{\lt}{\left(}
\newcommand{\rt}{\right)}
\newcommand{\lqu}{\left[}
\newcommand{\rqu}{\right]}

\newcommand{\diz}{(z)}
\newcommand{\dizpr}{(z')}
\newcommand{\dit}{(t)}
\newcommand{\mincir}{\raise-3.truept\hbox{\rlap{\hbox{\hspace{3.truept}
$\sim$}}\raise4.truept\hbox{\hspace{3.truept}$<$}\ }}
\newcommand{\magcir}{\raise-3.truept\hbox{\rlap{\hbox{\hspace{3.truept}
$\sim$}}\raise4.truept\hbox{\hspace{3.truept}$>$}\ }}
\newcommand{\minmag}{\raise-2.truept\hbox{\rlap{\hbox{$<$}}
\raise6.truept\hbox{$>$}\ }}
\newcommand{\htagl}{\hbox{\rlap{\hbox{$h$}}\raise0truept
\hbox{\hspace{3.truept}$\bar{ }$}\ }}

\begin{document}
\twocolumn[%

\title{Constraints on extended quintessence from 
high-redshift Supernovae}

\author{P. Caresia\altaffilmark{1}, S. Matarrese\altaffilmark{2} and
L. Moscardini\altaffilmark{3}}

\begin{abstract}
We obtain constraints on quintessence models from magnitude-redshift
measurements of 176 type Ia Supernovae. The considered quintessence
models are ordinary quintessence, with Ratra-Peebles and SUGRA
potentials, and extended quintessence with a Ratra-Peebles
potential. We compute confidence regions in the $\Omega_{m0}-\alpha$
plane and find that for SUGRA potentials it is not possible to obtain
useful constraints on these parameters; for the Ratra-Peebles case,
both for the extended and ordinary quintessence we find $\alpha\mincir
0.8$, at the $1\sigma$ level. We also consider simulated dataset for
the SNAP satellite for the same models: again, for a SUGRA potential
it will not be possible to obtain constraints on $\alpha$, while with
a Ratra-Peebles potential its value will be determined with an error
$\mincir 0.6$. We evaluate the inaccuracy made by approximating the
time evolution of the equation of state with a linear or constant
$w\diz$, instead of using its exact redshift evolution. Finally we
discuss the effects of different systematic errors in the
determination of quintessence parameters.

\end{abstract}

] 
\altaffiltext{1}{Dipartimento di Astronomia, 
  Universit\`a di Padova, vicolo dell'Osservatorio 2,
  I-35122 Padova, Italy; caresia@pd.astro.it}
\altaffiltext{2}{Dipartimento di Fisica `Galileo Galilei',
  Universit\`a di Padova, and INFN, Sezione di Padova, via Marzolo 8,
  I-35131 Padova, Italy; matarrese@pd.infn.it}
\altaffiltext{2}{Dipartimento di Astronomia, 
  Universit\`a di Bologna, via Ranzani 1,
  I-40127 Bologna, Italy; moscardini@bo.astro.it}

\section{Introduction}\label{sec1}

Cosmological tests such as the baryon fraction in galaxy clusters
(Hradecky et al. 2000; Allen et al. 2002), the abundance of massive
galaxy clusters (see e.g. Bahcall 2000; Borgani et al. 2001), the
magnitude-redshift relation for type Ia Supernovae (Riess et al. 1998;
Perlmutter et al. 1999; Gott et al. 2001; Tonry et al. 2003), the
statistical analysis of the galaxy distribution in large redshift
catalogs (e.g. Percival et al. 2001; Verde et al. 2002) and Cosmic
Microwave Background (CMB) anisotropies \cite{benn2003,sper2003}
indicate a low value for the matter density parameter today,
$\Omega_{m0}$, probably lying in the interval $0.15\mincir \Omega_{m0}
\mincir 0.4$.  Recent results from the study of CMB anisotropies
obtained with the {\it WMAP} satellite (Bennett et al. 2003) also
provide strong support for a flat (or very nearly flat) Universe.
Combining these two different indications leads to the hypothesis that
a new form of energy, named {\it dark energy} (DE), fills the gap
between $\Omega_{m0}$ and unity: $\Omega_{m0}+\Omega_{DE0}=1$.

One of the main goals of modern cosmology is to explain the nature of
dark energy. The simplest solution to this problem is to introduce a
cosmological constant $\Lambda$ in our Universe model. This scenario
can have a simple theoretical interpretation, as $\Lambda$ can be
related to the energy density of the vacuum state in quantum field
theory (Zel'dovich 1968).  Unfortunately, this simple explanation
results in a very large (formally infinite) value for the vacuum
energy density, which is larger by tens of orders of magnitude than
the observed one (of the order of $10^{-47}$ GeV$^4$). What emerges is
a fine-tuning problem, namely the ``cosmological constant
problem''. Another apparently unnatural feature of the cosmological
constant model is that it starts to dominate the Universe evolution
only in the very near past. This issue is usually referred to as the
``cosmic coincidence problem'' and reduces to a fine-tuning problem in
the choice of the initial conditions, in particular of the value of
$\Lambda$.

A possible way of alleviating these problems is to allow for a time
variation of the dark energy density, which is constant if due to
$\Lambda$. A very interesting class of models with this property goes
under the name of {\it quintessence} (Coble et al. 1997; Caldwell et
al. 1998; Ferreira \& Joyce 1998; Viana \& Liddle 1998).

One of the most promising cosmological tests on the properties of the
dark energy component is based on the already mentioned
magnitude-redshift relation for type Ia Supernovae (see e.g. Brax \&
Martin 1999; Podariu \& Ratra 2000; Podariu et al. 2001; Goliath et
al. 2001; Weller \& Albrecht 2001, 2002; Eriksson \& Amanullah 2002;
Gerke \& Efstathiou 2002; Padmanabhan \& Choudhury 2003; Di Pietro \&
Claeskens 2003; Knop et al. 2003).  In fact, these objects can be
considered as good standard candles, which makes it possible to
determine their luminosity distance, whose dependence on redshift is
specific of each particular cosmological framework.

In this paper we will focus on three different kinds of quintessence
models whose features are briefly described in Section
\ref{sec2}. In Section \ref{sec3} we present the constraints
obtained on these models using the magnitude-redshift relation for
existing type Ia Supernovae (SNIa) data. In Section \ref{sec4} we
illustrate how the constraints will improve with the SuperNovae
Acceleration Probe (SNAP) satellite which is currently being projected
and will be devoted to the discovery and study of SNIa (Aldering et
al. 2002; see also: http://snap.lbl.gov); we also check the validity
of approximating the exact redshift evolution of the equation of state
with a linear or constant behavior and discuss the possible effects of
different systematic errors on the parameter determination.  Finally,
in Section \ref{sec5} we present our conclusions.

\section{Theoretical framework}\label{sec2}

In this paper we focus on the Extended Quintessence (EQ) models,
introduced in (Perrotta et al. 1999; Baccigalupi et
al. 2000). Extended quintessence and related models have been also
considered in (Chiba 1999, 2001; Uzan 1999; Bartolo \& Pietroni 2000;
Boisseau et al. 2000; de Ritis et al. 2000; Faraoni 2000; Fujii 2000;
Chen et al. 2001; Bean 2001; Gasperini 2001; Perrotta \& Baccigalupi
2002; Riazuelo \& Uzan 2002; Torres 2002; Kneller \& Steigman 2003).

For these models the evolution of the scale factor $a$ and the scalar
field $\phi$ responsible for the quintessence component can be
obtained by solving the set of equations:
\begin{equation}\label{eq1}
{H}^2\equiv \lt\frac{\dot{a}}{a}\rt^2=\frac{1}{3F}\lt
a^2\rho_{fluid}+\frac{1}{2}\dot{\phi}^2+a^2V-3{H}\dot{F}\rt,
\end{equation}
\begin{equation}\label{eq2}
\ddot{\phi}+2{H}\dot{\phi}=\frac{a^2}{2}F_\phi R-a^2V_\phi\ .
\end{equation}
In the previous equations the dots denote derivatives with respect to
the conformal time and the subscript $\phi$ denotes differentiation
with respect to the scalar field; $R$ is the Ricci scalar;
$\rho_{fluid}$ represents the energy density associated with all the
constituents of the Universe except for the quintessence scalar field;
$V\lt\phi\rt$ is the quintessence potential and finally $F\lt\phi\rt$
is a function specifying the form of the coupling between $\phi$ and
gravity. Hereafter we will always refer to the non-minimally coupled
(NMC) scalar field models \cite{perr1999}, for which the function $F$
is defined as $F\lt\phi\rt={1}/{8\pi
G}+\tilde{F}\lt\phi\rt-\tilde{F}\lt\phi_0\rt$, where
$\tilde{F}\lt\phi\rt=\xi\phi^2$. This kind of models has two free
parameters: the dimensionless constant $\xi$, parametrizing the amount
of coupling, and the present value of the scalar field $\phi_0$.

The coupling between the scalar field and gravity generates a
time-varying gravitational constant (see e.g. the review by Uzan
2003). Upper bounds on this variation come from local laboratory and
solar system experiments \cite{gill1997} and from the effects induced
on photon trajectories (Reasenberg et al. 1979; Will 1984; Damour
1998).  As pointed out by Perrotta et al.  (2000), these bounds become
constraints on the parameters of the models:
\begin{equation}
\label{bounds}
32 \pi  G  \xi^2  \phi_0^2 \le \frac{1}{500}.
\end{equation}

We will use the previous inequality in the next sections in order to
improve our determination of the cosmological parameters.

For completeness and in order to allow a comparison with similar
analyses, we will also consider the case of ordinary quintessence
(OQ), i.e. models for which there is no direct coupling between the
scalar field and gravity (it is often referred to as the minimal
coupling case). OQ can be easily obtained from EQ in the limit of
$\xi\rightarrow0$.

If we want to completely specify a quintessence model, we have to
choose the analytical form for the potential $V\lt\phi\rt$. One of the
main advantages of a time-varying dark energy density is that it is
possible to alleviate the cosmic coincidence problem. This is achieved
by assuming particular classes of potentials, the so-called ``tracker
potentials'' (Steinhardt et al. 1999), for which one obtains at low
redshifts the same time evolution for $a$ and $\phi$, even starting
from initial conditions which differ by orders of magnitude: this
leads to a dark energy dominated era close to the present time,
without fine-tuning on the initial conditions.  For the following
analysis, we will consider two different classes of tracker
potentials: the inverse power-law Ratra-Peebles potential (hereafter
RP; Ratra \& Peebles 1988; see also Peebles \& Ratra 2003 and
references therein)
\begin{equation}\label{eq3}
V\lt\phi\rt=\frac{M^{4+\alpha}}{\phi^\alpha},
\end{equation}
and the SUGRA potential (Bin\inve truy 1999; Brax \& Martin 1999):
\begin{equation}\label{eq4}
V=\frac{M^{4+\alpha}}{\phi^\alpha}\exp{\lt4\pi G\phi^2\rt}.
\end{equation}
In particular, we will use the potential (\ref{eq3}) in the context of
both EQ and OQ models, while the potential (\ref{eq4}) will be
considered in the minimal coupling ($\xi=0$) case, only.

We solved numerically Eqs.(\ref{eq1}) and (\ref{eq2}) in the tracking
regime. The behaviors we found for $a(t)$ and $\phi(t)$ (not shown in
figure) are in excellent agreement with those obtained by Baccigalupi
et al. (2000), whose analysis also includes the Ratra-Peebles case in
the framework of EQ. In the next section we will use these results to
obtain our theoretical estimates of the luminosity distance $d_L$.

\section{Constraints from present high-redshift supernovae data}\label{sec3}
 
The purpose of this section is to test the possibility of constraining
the cosmological parameters describing quintessence models by using
the best SNIa dataset presently available.  To this aim we build a
sample combining data coming from the literature. As a starting point,
we consider the data reported in Table 15 of Tonry et al. (2003).  In
particular, we use for our analysis the subset which is presented by
the authors as the most reliable one for cosmological studies. This
data compilation, comprising 172 SNIa, is obtained by eliminating from
the original whole sample of 230 SNIa the objects at low redshift ($z
< 0.01$) and those with high reddening ($A_V> 0.5$ mag).  Then, we
consider the data from Table 3 of Knop et al. (2003), but including in
the sample only SNIa belonging to their ``low-extinction primary
subset'' (7 objects). In their Table 4, Knop et al. (2003) present the
new fits to the Perlmutter et al. (1999) data, which are already
included in the Tonry et al. (2003) sample. In this case we decided to
use the magnitudes from Knop et al. (2003), because they are obtained
using new fitting lightcurves. The two catalogues have also SNIa in
common in the low-redshift sample, taken from Hamuy et al. (1996) and
Riess et al. (1999). For these we prefer to use the data reported by
Tonry et al. (2003) because they are obtained as a median of different
fitting methods.  For coherence with our previous choice, we decided
to exclude 5 objects, present in the low-redshift and Perlmutter et
al. (1999) samples, but excluded by the ``low-extinction primary
subset'' of Knop et al. (2003). Finally, we add two new SNIa, 2002dc
and 2002dd, recently studied by Blakeslee et al. (2003). Therefore the
sample of SNIa here considered comprises 176 objects.

To constrain the cosmological parameters, we compare through a
$\chi^2$ analysis the redshift dependence of the observational
estimates of $\log{\lt d_L\rt}$ to their theoretical values, which for
a flat, matter or dark energy dominated universe can be obtained as
\begin{equation}\label{eq14}
d_L\diz =\frac{1+z}{H_0} \int_0^z{\lqu
\frac{\Omega_{m0}\lt1+z'\rt^3+\Omega_{\phi0}f\dizpr}
{8\pi G F}
\rqu^{-{1}/{2}}dz'},
\end{equation}
where
\begin{equation}\label{eq15}
f\diz=\exp{\lt3\int_0^z{\frac{1+w\dizpr}{1+z'}dz'}\rt}.
\end{equation}
Here $H_0$ is the present value of the Hubble constant, and
$\Omega_{m0}$ and $\Omega_{\phi0}$ represent the contributions to the
present density parameter due to matter and scalar field,
respectively. The quantity $w\equiv p_\phi/\rho_\phi$ sets the
equation of state relating the scalar field pressure
\begin{equation}
p_\phi=\frac{\dot{\phi}^2}{2 a^2}+V(\phi)-\frac{3 H \dot{F}}{a^2}
\end{equation}
and its energy density 
\begin{equation}
\rho_\phi=\frac{\dot{\phi}^2}{2 a^2}-V(\phi)
+\frac{\ddot{F}}{a^2}+\frac{H \dot{F}}{a^2}\ .
\end{equation}
The quantity $w(z)$ can be computed by using the numerical solutions
of the previous equations, once a cosmological model is assumed.

We can now find, for the different quintessence models described in
the previous section, the parameters (and their confidence regions)
which best fit the SNIa data by using the standard $\chi^2$ method.
For this analysis we have to consider the errors on the distance
moduli. For the objects coming from Tonry et al. (2003), they are
obtained directly from their Table 15, once a value for $H_0$ is
assumed. On the other hand, Knop et al. (2003) report errors for the
apparent magnitudes of their SNIa. We obtain the uncertainty on the
distance modulus by adding in quadrature an error of 0.05 in the
estimate of the absolute magnitude, as suggested by the data of Hamuy
et al. (1996), Riess et al. (1999) and Knop et al. (2003), Table 4.
In addition, we include an extra contribution to the error coming from
the possible uncertainty in the peculiar velocities. In particular,
following Tonry et al. (2003) we add $500$ km s$^{-1}$ divided by the
redshift in quadrature to the distance error.

\begin{figure}[t]
\begin{center}
\includegraphics[height=0.45\textheight,width=0.5\textwidth]{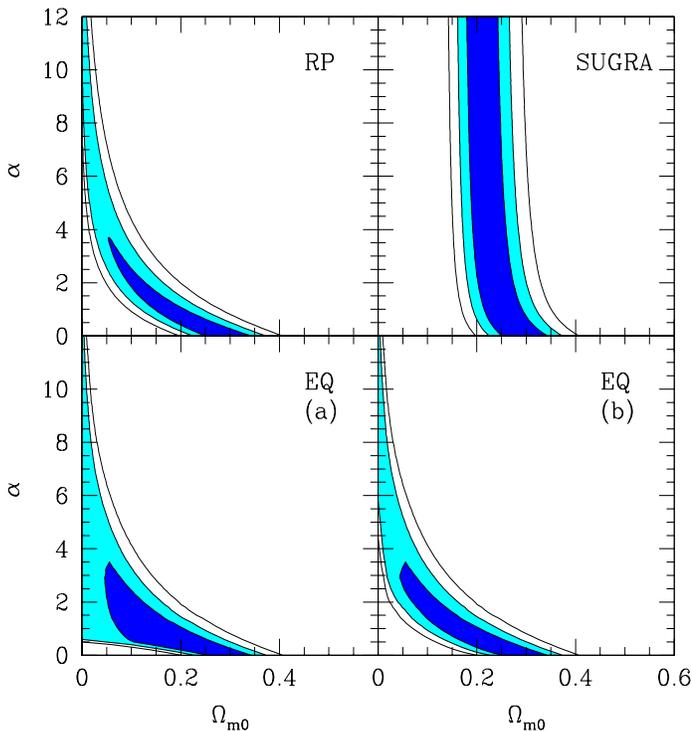}
\caption{
Confidence regions ($\Delta\chi^2=2.30, 5.99, 11.8$, corresponding to
1, 2 and 3$\sigma$ for a Gaussian distribution with two free
parameters, respectively) for the parameters $\Omega_{m0}$ and
$\alpha$, as obtained using the sample of 176 SNIa data presently
available.  Different panels refer to different quintessence models:
ordinary quintessence with a Ratra-Peebles potential (top left);
ordinary quintessence with a SUGRA potential (top right); extended
quintessence with a Ratra-Peebles potential (bottom left); extended
quintessence with a Ratra-Peebles potential when the upper limits on
the time variation of the gravitational constant are satisfied (bottom
right). }
\label{fig:sn}    
\end{center}
\end{figure}

We compute the values for $\chi^2$ on a regular grid of the considered
parameters ($\Omega_{m0}$, $\alpha$ and $\xi$ for the EQ models;
$\Omega_{m0}$ and $\alpha$ for the OQ models), looking for its minimum
value, $\chi^2_{\rm min}$. In particular, we allow $\Omega_{m0}$ to
vary between 0 and 1 with spacing of 0.001; $\alpha$ between 0 and 12
with spacing of 0.1 and $\xi$ between 0.001 to 0.100 with spacing of
0.001.  The results in the plane $\Omega_{m0}-\alpha$ are shown in
Figure \ref{fig:sn}, where we display the contours of constant
$\Delta\chi^2\equiv \chi^2-\chi^2_{\rm min}= 2.30, 5.99$ and 11.80,
corresponding to $1\sigma$, $2\sigma$ and $3\sigma$ for a Gaussian
distribution with two free parameters, respectively.

The top-left panel refers to the ordinary quintessence model with a RP
potential. In this case we find $\chi^2_{\rm min}=206$ for
$\Omega_{m0}=0.30$ and $\alpha=0.0$. However, from the plot it is
evident that there is a strong degeneracy between the two free
parameters $\alpha$ and $\Omega_{m0}$ and it is not possible to obtain
strong constraints on them at the same time.  Nevertheless, we can
extract some information by considering the $\chi^2$ distribution when
only a single parameter is allowed to vary. In this case, we obtain
the errorbars associated to the best fit value by assuming
$\Delta\chi^2=1$, which corresponds to $1\sigma$ for a Gaussian
distribution with one single parameter. The SNIa dataset we used does
not allow to obtain tight constraints on the parameter $\alpha$:
$\alpha<0.83$ at the $1\sigma$ confidence level, with a best fit value
of $\alpha=0$. In addition, we obtain
$\Omega_{m0}=0.30^{+0.03}_{-0.10}$, for the matter density
parameter. If we impose the Gaussian prior $\Omega_{m0}=0.27\pm0.04$,
as suggested by the combined analysis of recent CMB observations and
large-scale structure data \cite{sper2003}, we obtain $\alpha<0.47$ at
the $1\sigma$ confidence level (best fit value: $\alpha=0.10$). We
notice that our results are in agreement with those obtained from a
similar analysis carried out by Podariu \& Ratra (2000), even though,
thanks to the improvement in the SNIa dataset, the confidence contours
start to close off, at least at the $1\sigma$ confidence level.

In the top-right panel we report the results still for the OQ model,
but with the SUGRA potential. The values for $\chi^2_{\rm min}$ and
the corresponding parameters are the same obtained for the RP
potential.  In this case the confidence regions are almost vertical,
showing a very small dependence on $\alpha$: this does not allow us to
extract constraints on $\Omega_{m0}$ and $\alpha$.  Only if we impose
the Gaussian prior $\Omega_{m0}=0.27\pm0.04$, we are able to obtain
$\alpha < 2.78$ at the $1\sigma$ level (best fit value:
$\alpha=0.24$).

In the two bottom panels of Figure \ref{fig:sn} we present the results
for the model with EQ and a RP potential. In this case we have three
free parameters: in addition to $\alpha$ and $\Omega_{m0}$, there is
$\xi$, which parametrizes the strength of the coupling between the
scalar field and gravity.  In the bottom-left panel we show the
two-dimensional confidence regions in the $\alpha-\Omega_{m0}$ plane
as obtained by minimizing $\chi^2$ with respect to $\xi$: they appear
very similar to the previous case of ordinary quintessence with RP
potential, only slightly larger. Again it is convenient to discuss the
results when a single free parameter is considered.  Unfortunately,
there are no possibilities to obtain constraints on $\xi$ (best fit
for $\xi=0.001$), even at the $1\sigma$ confidence level.  For the
other two parameters our results are again very similar to that of
ordinary quintessence with RP potential: $\alpha<0.82$ (best fit
$\alpha=0.02$) and $\Omega_{m0}=0.28^{+0.05}_{-0.09}$.  The
bottom-right panel shows how the confidence regions change when the
upper limits on the time variation of the gravitational constant are
taken into account. The constraints on $\alpha$ and $\Omega_{m0}$ do
not change significantly: $\alpha<0.82$ and
$\Omega_{m0}=0.28^{+0.04}_{-0.09}$, respectively.  Considering $\xi$,
the inclusion of the upper bounds does not improve the situation, and
$\xi$ remains completely undetermined.

Finally, instead of imposing the constraints just described, we can
consider the prior $\Omega_{m0}=0.27\pm0.04$. Again, we cannot obtain
useful bounds on $\xi$, while we obtain $\alpha<0.45$ at the $1\sigma$
level (best fit: $\alpha=0.11$).  The previous results clearly show
the necessity for new samples of SNIa. The SNAP satellite will be
dedicated to this purpose: in the next section we will check the
improvement that it will allow on the determination of the
cosmological parameters.

The results presented in this section assume that all the data points
are purely governed by statistical errors. Even if we carefully
selected our SNIa sample in order to reduce possible systematic
errors, their importance in the present data cannot be completely
excluded.  As largely discussed in Knop et al. (2003, see also Tonry
et al. 2003), systematics can have different origins, In particular
Knop et al. (2003) studied the effects on the determination of the
cosmological parameters (they considered $\Omega_{m0}$,
$\Omega_{\phi0}$, $w$) coming from different SNIa selection,
lightcurve fitting methods, contamination from non-type Ia supernovae,
Malmquist bias, gravitational lensing, dust properties, etc. Their
conclusion is that the total systematic error is of the same order of
magnitude as the statistical uncertainty.  Because of the weakness of
the constraints we obtained on $\alpha$ and $\xi$ using present SNIa
data, we prefer to leave an extended discussion of systematics to the
next section, where we will present the expected results from the SNAP
satellite.  However, here we report the results of some tests which
have been done.  In order to verify if our sample suffers from
selection effects, we repeated our analysis by considering only 54
SNIa belonging to the low-extinction primary subset of Knop et
al. (2003) and coming only from the Supernova Cosmology
Project. Because of the smallness of this sample, the resulting
constraints cannot be directly compared to those obtained from the
whole sample of 176 SNIa, being weaker, but we notice that the
best-fit values for $\Omega_{m0}$ and $\alpha$ are only slightly
different.  Then, we checked the systematic effect due to the possible
type-contamination, by excluding from our sample those objects whose
confirmations as type Ia supernovae are questionable. This new
subsample has a smaller number of very-high redshift SNIa, and
consequently the constraint on $\alpha$ becomes much weaker when
compared with the whole sample: $\alpha<1.62$ for OQ with RP
potential, and $\alpha<1.52$ for EQ, always with RP
potential. Finally, we checked that the exclusion of very-low redshift
objects (z<0.03), whose measurements of distance moduli are possibly
affected by peculiar velocities, does not change the resulting
confidence levels, confirming that our treatment of this kind of error
is reliable and that the constraining power of the analysis comes from
high-redshift SNIa.


\section{Constraints from future high-redshift supernovae data}\label{sec4}

In order to improve the sample of studied SNIa, a new satellite, the
SuperNovae Acceleration Probe (SNAP) is currently under project.  SNAP
is expected to perform in two years a complete spectroscopic and
photometric analysis for approximately 2000 high-redshift SNIa
reaching a maximum redshift of $z=1.7$. Due to the large increase of
both the number of observed objects and the covered redshift interval,
SNAP can prove extremely useful for cosmological studies.

\begin{figure}[t]
\begin{center}
\includegraphics[height=0.35\textheight,width=0.4\textwidth]{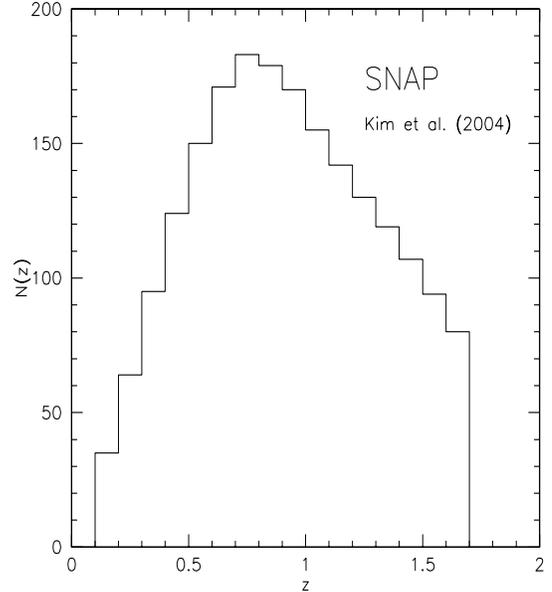}
\caption{The SNIa redshift distribution used in the simulated SNAP samples
(see also Kim et al. 2004). }
\label{fig:hist}    
\end{center}
\end{figure}

In order to check the ability of this satellite to constrain the
parameters for quintessence models, we generate pairs of distance
moduli and redshift for 1998 SNIa, adopting the fiducial distribution
proposed by Kim et al. (2004) and shown in Figure
\ref{fig:hist}. Always as in Kim et al. (2004),  
we include in addition a very low-redshift sample of 300 supernovae
with redshift uniformly distributed between $z=0.03$ and
$z=0.08$. Thus, the total number of objects considered in the
following analysis is 2298.

The simulated distance moduli $\mu$ are assumed to have a Gaussian
distribution around the true value, with a dispersion
$\sigma_\mu=0.15$. The true distance modulus for a given redshift $z$
is known once we specify the background cosmology.  For our analysis,
we decided to simulate SNIa datasets with three different background
cosmologies, whose properties and parameter values are reported in
Table \ref{tab2}. For all datasets, we use for the Hubble constant
$H_0=71$ km s$^{-1}$ Mpc$^{-1}$, which is the value suggested by the
combined analysis of WMAP and large-scale structure data
\cite{sper2003}.

\begin{table}[h]
\caption{Properties and values of the parameters for the
different cosmological models assumed in the generation of simulated
SNAP samples.}
\begin{center}
\begin{tabular}{cccccc}
\hline
Name & Model & Potential & $\alpha$ & $\Omega_{m0}$ & $\xi$ \\
\hline
dataset 1 & OQ & RP & $1.0$ & $0.3$ 
& $0.000$ \\
dataset 2 & OQ & SUGRA  & $1.0$ & $0.3$ 
& $0.000$ \\
dataset 3 & EQ & RP  & $1.0$ & $0.3$ 
& $0.015$ \\
\hline
\end{tabular}
\label{tab2}
\end{center}
\end{table}

Adopting the same $\chi^2$ method described in the previous section,
we then analyze these simulated datasets, fitting each quintessence
model only with the data obtained from the same background cosmology.
In computing the value for $\chi^2$ we consider an error of 0.15 mag
on the distance moduli and we neglect the errors on the redshifts $z$.

It is also quite important to verify how good is the approximation of
using a linear (i.e. $w\diz=w_0 + w_1 z$, with $w_0$ and $w_1$
suitable constants) or a constant ($w=w_0={\rm const}.$) equation of
state in the fitting procedure (as often done in the literature),
rather than its exact redshift evolution (as we did in the previous
section).  To this purpose we apply the following procedure. Let us
refer, for simplicity, to one of the OQ models.  For each pair of
values $\lt\Omega_{m0},\alpha\rt$, we numerically obtain the redshift
evolution of the equation of state computing $w\diz$ in 1700 equally
spaced values of $z$ in the interval $\lqu 0,1.7\rqu$.  Using the
Least squares method, we determine the straight line which best fits
this ensemble of points. Finally, we use the theoretical distance
modulus obtained from this linear approximation for $w$ instead of its
exact evolution to calculate the value of $\chi^2$ with the SNIa
simulated data.  Similarly, in order to check the validity of the
approximation of a constant equation of state, we follow the same
procedure but substituting the straight line with the mean value of
$w\diz$ in the considered range. In the following figures we will
illustrate the differences in the confidence regions between these two
approximations and the results obtained assuming the exact equation of
state.

\begin{figure}[t] 
\begin{center} 
\includegraphics[height=0.45\textheight,width=0.5\textwidth]
{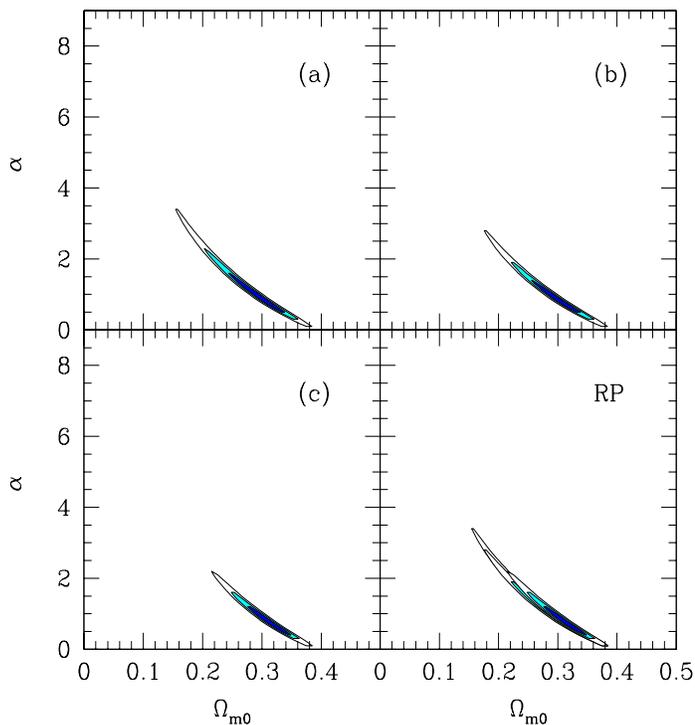} 
\caption{
Confidence regions ($\Delta\chi^2=2.30, 5.99, 11.8$, corresponding to
1, 2 and 3$\sigma$ for a Gaussian distribution with two free
parameters, respectively) for the parameters $\Omega_{m0}$ and
$\alpha$ for the ordinary quintessence model with a Ratra-Peebles
potential. The results are obtained with the simulated SNAP sample
(dataset 1).  Panel (a) refers to the constraints obtained by assuming
the exact equation of state $w\diz$; panel (b) to the linear $w\diz$
approximation; panel (c) to the constant $w$ approximation. The final
panel shows the superposition of the three cases.}
\label{fig:snapratra} 
\end{center}
\end{figure}

In Figure \ref{fig:snapratra} we show the confidence regions in the
$\alpha-\Omega_{m0}$ plane obtained by fitting the OQ model with a RP
potential to the dataset 1.  As done also for the other fits described
in this section, we fix the value of the Hubble constant to the `true'
value, $H_0=71$ km s$^{-1}$ Mpc$^{-1}$.  In the panel (a) we show the
results obtained with the exact equation of state. The best-fitting
parameters are: $\alpha=0.91^{+0.44}_{-0.33}$ and
$\Omega_{m0}=0.30^{+0.03}_{-0.04}$ at the $1\sigma$ level. They are
consistent with the values assumed in the background cosmology, within
the error limits.  By comparing these results to those presented in
the top-left panel of Figure \ref{fig:sn}, it is also evident the
improvement with respect to the constraints obtained with the present
SNIa data: the confidence regions are much narrower. Nonetheless,
there is still a little degeneracy between $\alpha$ and $\Omega_{m0}$:
to solve it will be useful to combine magnitude-redshift measurements
to other cosmological tests (see, e.g., Balbi et al. 2003; Frieman et
al. 2003; Jimenez et al. 2003; Caldwell \& Doran 2003). The case of OQ
model with a RP potential has already been examined by Podariu et
al. (2001) for SNAP simulated data. Their confidence regions in the
$\Omega_{m0}-\alpha$ plane look quite similar to ours, even if they
considered a fiducial cosmology with $\alpha=4$ and $\Omega_{m0}=0.2$.
A similar analysis has been carried out also by Eriksson \& Amanullah
(2002), who found an error on $\alpha$ which is approximately twice
larger than ours. This difference is probably due to the different
redshift distribution adopted.

In panel (b) of Figure \ref{fig:snapratra} we show the confidence
regions obtained when $\chi^2$ is computed using the linear
approximation for $w\diz$, while in panel (c) we adopt a constant
equation of state.  The differences between the three considered cases
(exact, linear and constant $w$) are evident also from the last panel,
showing the superposition of the corresponding confidence regions.
The best-fitting values we obtain are $\alpha=0.81^{+0.37}_{-0.29}$
and $\Omega_{m0}=0.31^{+0.03}_{-0.03}$ for the linear $w\diz$, and
$\alpha=0.76^{+0.30}_{-0.26}$ and $\Omega_{m0}=0.31^{+0.02}_{-0.03}$
for the constant $w$, in both cases at $1\sigma$ level.  The measured
cosmological parameters are then consistent with the true values
within the error limits.  However, there seems to be a systematic
tendency to underestimate $\alpha$, which increases when we decrease
the accuracy used to describe the redshift evolution of
$w$. Nevertheless, this systematic offset will be too small to be put
in evidence by the SNAP data, even at the $1\sigma$ level.  On the
contrary, the approximations on $w\diz$ do not strongly influence the
determination of $\Omega_{m0}$. Finally, we notice that, at the
$1\sigma$ confidence level, the size of the errors for both $\alpha$
and $\Omega_{m0}$ appear to be smaller, i.e. systematically
underestimated, when the linear and constant approximations are used
instead of the exact solution.

\begin{figure}[t] 
\begin{center} 
\includegraphics[height=0.45\textheight,width=0.5\textwidth]
{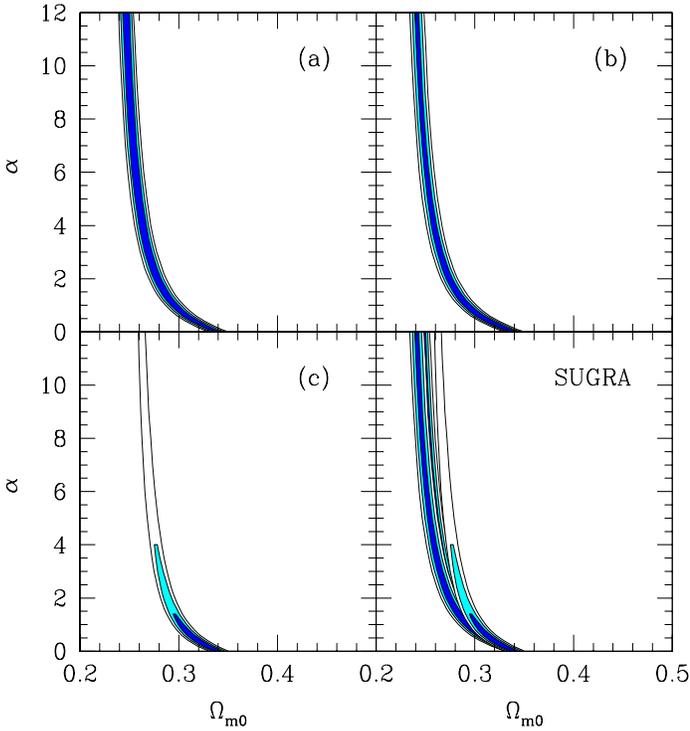}
\caption{
As Figure \ref{fig:snapratra}, but for the ordinary quintessence model
with a SUGRA potential. The results are obtained with the 
simulated SNAP sample (dataset 2). }
\label{fig:snapsugra} 
\end{center}
\end{figure}

We then perform the same kind of analysis by fitting OQ models with a
SUGRA potential using the dataset 2. The results are shown in Figure
\ref{fig:snapsugra}. Looking at panel (a), which refers to the results 
obtained with the exact $w\diz$, it is evident that the SUGRA models
present a strong degeneracy in $\alpha$, becoming larger with
increasing values of $\alpha$. This was already pointed out by Brax \&
Martin (1999), who studied the dependence of $w_0$ on $\alpha$ in the
same class of models. As a consequence, in this case it is impossible
to determine the errors on the cosmological parameters.  The
confidence regions under the approximation of linear equation of state
are shown in panel (b): they appear very similar to those displayed in
panel (a). Instead, if we assume a constant $w$ (see panel c), the
$1\sigma$ contour closes off, and we obtain:
$\alpha=0.34^{+0.59}_{-0.33}$ and $\Omega_{m0}=0.32\pm0.02$, which are
marginally consistent with the true values $\alpha=1$ and
$\Omega_{m0}=0.30$. Again, the last panel shows the superposition of
the three cases.  The situation slightly improves if we impose a
Gaussian prior on $\Omega_{m0}$, namely $\Omega_{m0}=0.30\pm0.05$.  In
fact we obtain $\alpha=0.95^{+6.25}_{-0.78}$ for the exact $w\diz$,
$\alpha=0.49^{+1.23}_{-0.40}$ for the linear $w\diz$ and
$\alpha=0.38^{+0.59}_{-0.26}$ for the constant $w$, all at the
$1\sigma$ confidence level.  All these values are consistent within
the error limits with the original assumption of $\alpha=1$, but we
see that by using the exact $w\diz$ the determination is very
poor. Moreover, the use of linear or constant equation of state
approximations could lead to an artificially higher precision in the
determination of $\alpha$. For these reasons, we can conclude that
even with SNAP it will not be possible to constrain OQ models with a
SUGRA potential.

Finally, we use the dataset 3 to estimate the best-fitting parameters
for EQ models with a RP potential. To this aim we determine the values
of $\chi^2$ on a three-dimensional grid of values for the parameters
$\alpha,\Omega_{m0}$ and $\xi$. In particular, we consider 100 values
for $\xi$, regularly spaced between 0.001 to 0.100, as done in the
previous section with the presently available SNIa data.

\begin{figure}[t]
\begin{center} 
\includegraphics[height=0.45\textheight,width=0.5\textwidth]
{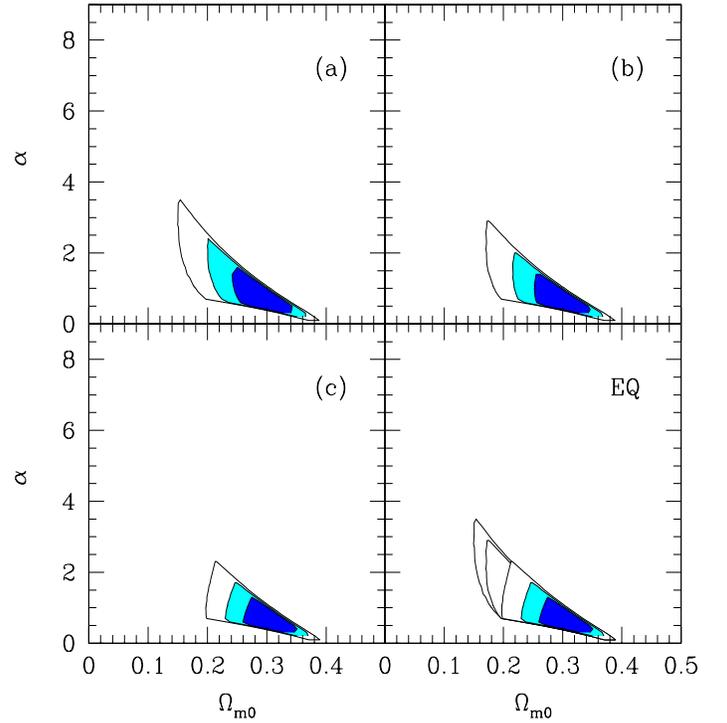} 
\caption{
As Figure \ref{fig:snapratra}, but for the extended quintessence model
with a Ratra-Peebles potential. The results are obtained with the
simulated SNAP sample (dataset 3). }
\label{fig:snapbid} 
\end{center}
\end{figure}

In Figure \ref{fig:snapbid} we show the confidence regions in the
$\Omega_{m0}-\alpha$ plane, obtained by minimizing (for each pair of
these parameters) $\chi^2$ with respect to $\xi$.  Panel (a) refers to
the exact $w\diz$ case and can be directly compared to the bottom-left
panel of Figure \ref{fig:sn}. Even if the regions become narrower,
they are still larger than in the OQ models.  It is possible to obtain
the one-dimensional distribution separately for each of the three
parameters, by minimizing $\chi^2$ with respect to the others. In this
way we find that there is very little dependence of the minimized
$\chi^2$ on $\xi$. Then, it is not possible to obtain useful
constraints on this parameter (best fit value: $\xi=0.001$). Instead,
we have $\alpha=0.95^{+0.44}_{-0.62}$ and
$\Omega_{m0}=0.30^{+0.02}_{-0.03}$ at the $1\sigma$ level. This values
are consistent with the true values ($\alpha=1$ and $\Omega_{m0}=0.3$)
within the errors.

The confidence regions obtained using the approximated (linear and
constant) equations of state (panels (b) and (c), respectively) are
quite different from the previous ones, as it is evident from the
superposition in the last panel.  Again, there is no possibility to
constrain $\xi$ (best fit values: $\xi=0.001$ for linear $w\diz$, and
$\xi=0.003$ for constant $w$).  For the remaining parameters we
obtain: $\alpha=0.84^{+0.37}_{-0.51}$ and
$\Omega_{m0}=0.31^{+0.02}_{-0.03}$ for the linear $w\diz$; and
$\alpha=0.78^{+0.31}_{-0.46}$ and $\Omega_{m0}=0.32^{+0.01}_{-0.03}$
for constant $w$, at $1\sigma$ confidence level. These values are
always consistent with the true ones, within the error limits.

\begin{figure}[t]
\begin{center} 
\includegraphics[height=0.45\textheight,width=0.5\textwidth]
{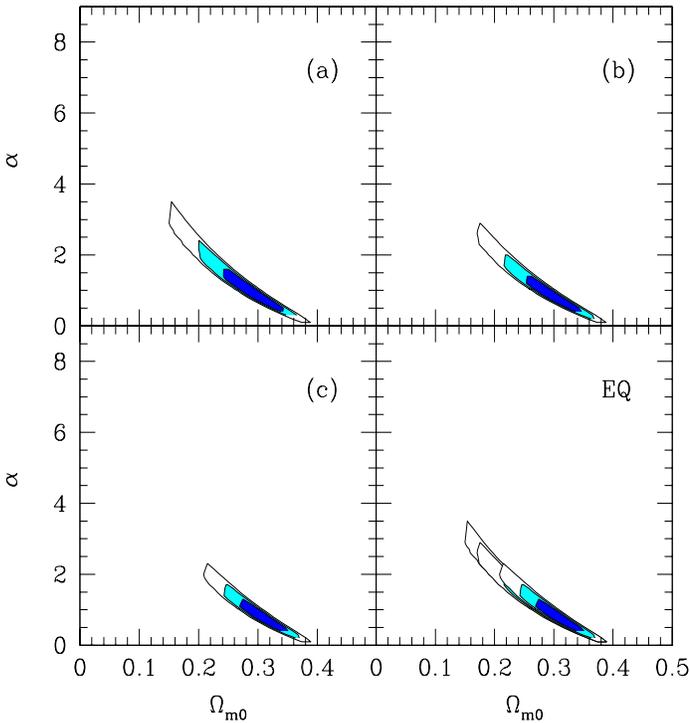} 
\caption{
As Figure \ref{fig:snapratra}, but for the extended quintessence model
with a Ratra-Peebles potential. The results are obtained by using the
simulated SNAP sample (dataset 3) and imposing the upper bound on the
time variation of the gravitational constant.}
\label{fig:snapvinc} 
\end{center}
\end{figure}

As a final point, it is interesting to discuss how the confidence
regions change if we impose the constraint coming from the time
variation of the gravitational constant. The results are shown in
Figure \ref{fig:snapvinc}, where we draw the confidence regions in the
$\Omega_{m0}-\alpha$ plane, obtained by minimizing $\chi^2$ with
respect to $\xi$ and considering only the combination of the
parameters for which the bounds on $G\dit$ are satisfied. This
condition excludes combination of the parameters for which $\alpha$
and $\xi$ are high, and this is the main motivation for the
differences with respect to Figure \ref{fig:snapbid}.  Considering the
case where the exact equation of state is used (panel (a)), we have
$\alpha=0.95\pm{0.43}$ and $\Omega_{m0}=0.30\pm{0.03}$ at $1\sigma$:
there is no significant change in the determination of these
parameters with the imposition of the constraint. However, this time
we are able to obtain an upper limit on $\xi$: $\xi<0.028$ at the
$1\sigma$ level. This result is consistent with the assumed true
value: $\xi=0.015$.  Panel (b) refers to the linear $w\diz$
approximation, for which we derive $\alpha=0.84^{+0.37}_{-0.38}$,
$\Omega_{m0}=0.31^{+0.02}_{-0.03}$ and $\xi<0.030$ at the $1\sigma$
level. Finally, panel (c) considers the constant $w$ approximation. In
this case the resulting constraints are $\alpha=0.78^{+0.32}_{-0.33}$,
$\Omega_{m0}=0.32^{+0.01}_{-0.03}$ and $\xi<0.030$ at the $1\sigma$
level. Again, all this values are consistent with the true ones.

Up to now, we discussed the possibility of determining the
quintessence parameters in the future data coming from SNAP by
assuming intrinsic statistical errors only. Here we consider the
impact of systematic errors on the previous results. Kim et al. (2004)
define two general forms for them: uncorrelated systematic
uncertainties and magnitude offsets.  The first case represents a
random dispersion which cannot be reduced below a given magnitude
error over a finite redshift bin (here assumed to be $\Delta z=0.1$):
possible examples are calibration errors and imperfect galaxy
subtraction. The second case is a coherent shift acting as a bias on
all SNIa magnitudes: selection effects as Malmquist bias or detector
problems can produce this kind of effect.

\begin{figure}[t]
\begin{center} 
\includegraphics[height=0.45\textheight,width=0.5\textwidth]
{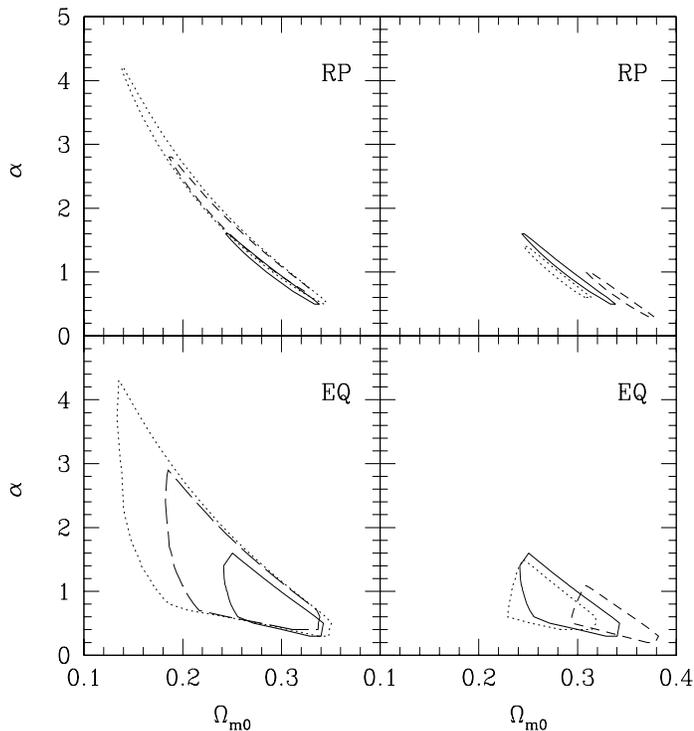}
\caption{Effects of systematics errors on the
1$\sigma$ confidence regions for the parameters $\Omega_{m0}$ and
$\alpha$ for the ordinary quintessence model with a Ratra-Peebles
potential (upper panels) and extended quintessence model with a
Ratra-Peebles potential (lower panels).  The results are obtained with
the simulated SNAP samples (datasets 1 and 3, respectively).  Left
panels show the effects of irreducible systematic uncertainties:
dotted (dashed) contours refer to confidence regions obtained by
including constant (linearly increasing) error $\Delta m$ (see text
for details).  Right panels present the effects of magnitude offsets:
dotted (dashed) contours refer to a positive (negative) shift of 0.02
mag.  In all panels, the solid curve represents the constraints
obtained by assuming the presence of intrinsic statistical errors,
only.}
\label{fig:systematics} 
\end{center}
\end{figure}

In order to simulate irreducible systematics, we strictly follow Kim
et al. (2004). In particular, for each redshift bin, containing
$N_{SN}$ objects (see Figure \ref{fig:hist}), we add in quadrature to
the intrinsic magnitude dispersion per SNIa (0.15 mag) an irreducible
magnitude error $\Delta m$:
\begin{equation}
\sigma_m=\sqrt{\frac{0.15^2}{N_{SN}}+(\Delta m)^2}\ .  
\end{equation} 
As in Kim et al. (2004), we consider two different possibilities for
$\Delta m$: a constant value equal to 0.02 mag, which is the target of
the SNAP mission, and a linear function increasing with redshift
reaching at the maximum covered redshift, $z=1.7$, the maximum
expected error for SNAP, 0.02 mag: $\Delta m=0.02(z/1.7)$. The
resulting constraints on $\Omega_{m0}$ and $\alpha$ are shown in
Figure \ref{fig:systematics} for the RP potential in the case of OQ
and EQ (top-left and bottom-left panels, respectively).  In both cases
we show the 1$\sigma$ confidence regions when no priors are
considered.  The solid contour refers to the constraints obtained by
considering statistical errors only; they are the same shown in the
top-left panel of Figures \ref{fig:snapratra} and \ref{fig:snapbid}.
Dotted and dashed lines present the 1$\sigma$ regions when we include
constant and linear irreducible systematic errors. We notice that
systematics, as expected, extend the confidence regions, in particular
in the direction of lower values for $ \Omega_{m0}$ and higher values
for $\alpha$.  This effect is larger when we apply a constant $\Delta
m$.  In the OQ case, the best-fitting values for $\alpha$ is shifted
up of 0.1 and the errors on the two variables are strongly correlated.

Then we investigate the effect of systematic magnitude offsets.
First, we consider a constant shift of 0.02 mag, both positive and
negative. The analysis (not shown in the figure) confirms the results
obtained by Kim et al. (2004): in this case the best-fitting
parameters are the true values with errors which are very similar to
the ones obtained by considering statistical errors only, without any
systematic biases.  Then we consider a magnitude offset which is
linearly proportional to the redshift as $\Delta m=\pm
0.03(z/1.7)$. The results for the same models previously considered
(OQ and EQ with RP potential) are shown in the right panels of Figure
\ref{fig:systematics}.  As discussed by Kim et al. (2004), an offset
in magnitude can give best-fitting parameters which are wrong with
smaller confidence regions, i.e.  it is possible to have very accurate
but wrong answers.  This is exactly what is happening in our
case. Considering a positive linear offset, we find the tendency to
have confidence regions systematically shifted towards the left, while
we have the opposite trend for negative magnitude shifts: the true
values of the parameters $\Omega_{m0}$ and $\alpha$ are excluded at
1$\sigma$ confidence level.  Finally we consider the presence of a
possible mismatch between the calibration of low- and high-redshift
SNIa. This is done by introducing a constant offset of $\pm0.02$ mag
to the 300 local objects only. The results (not shown in the figure
for clarity) show that this kind of effect produces a small
enlargement of the confidence regions, without introducing systematic
biases in the parameter determination.

\section{Conclusions}\label{sec5}

The main goal of this paper has been to discuss the possibility of
using SNIa distances and redshifts to obtain reliable constraints on
the parameters defining the quintessence models. In particular we
considered extended quintessence models with Ratra-Peebles potential,
and, for completeness, ordinary quintessence models with both
Ratra-Peebles and SUGRA potentials.

As a first step, we studied the constraints which result from the
analysis of the largest SNIa sample presently available (176 objects,
coming from Tonry et al. 2003, Knop et al. 2003 and Blakeslee et
al. 2003).  To this purpose, we used the exact redshift evolution of
the equation of state $w\diz$, as numerically determined, avoiding any
approximation.  Our results show that for ordinary quintessence with a
SUGRA potential it is not possible to obtain significant limits on the
potential exponent $\alpha$, because of the weak dependence of the
equation of state on it.  For ordinary quintessence with a
Ratra-Peebles potential, it is possible to obtain an upper limit on
$\alpha$: at the $1\sigma$ confidence level we find $\alpha\le 0.83$.
Our results, which are in agreement with a similar analysis made by
Podariu \& Ratra (2000) who used an older SNIa dataset, are consistent
with a cosmological constant model, for which $\alpha=0$.  We obtained
a similar constraint for $\alpha$ when considering the extended
quintessence models with Ratra-Peebles potential.  Unfortunately, it
is not possible to obtain useful constraints on the non-minimal
coupling parameter $\xi$ between the scalar field and gravity.

We then discussed the potential improvement on the previous results
when future SNIa samples, with a larger number of objects and more
extended redshift coverage, will be available.  To this purpose, we
simulated SNIa datasets in different cosmological models, with the
characteristics of the expected SNAP satellite observations (almost
2000 objects up to $z=1.7$).
 
For ordinary quintessence with a SUGRA potential, we found that it
will still be difficult to constrain the parameters even with the SNAP
SNIa data.  Considering models with the Ratra-Peebles potential, both
in extended and ordinary quintessence, our results suggest that
$\alpha$ can be determined with an error $\mincir 0.6$ (at the
$1\sigma$ significance level), while $\Omega_{m0}$ will be constrained
with an error of approximately 0.03 (always at $1\sigma$).  In
extended quintessence models, even by imposing the upper bounds on the
time variation of the gravitational constant, it will be possible to
obtain only an upper limit on the coupling constant: $\xi<0.028$.

As a final issue, we discussed the systematic errors on the parameter
estimates which originate if a constant or linear approximation is
used for the equation of state, $w\diz$, instead of the exact redshift
evolution. For all the considered quintessence models, the confidence
regions obtained with these approximations look narrower than the
exact ones. As a consequence, these approximations on $w\diz$ lead to
a systematic underestimate of the errors.  Nevertheless, the set of
cosmological parameters determined by the fitting procedure are
consistent with that used in the simulations.  This means that the
systematic errors induced by the assumed approximations are still
smaller than the precision allowed on the cosmological parameters by
SNAP, even at the $1\sigma$ confidence level.

\acknowledgements{This work has been partially supported by Italian MIUR 
and INAF.  PC thanks the Centre of Excellence MIUR ``Science and
Applications of Advanced Computing Paradigms'' for financial
support. We acknowledge Carlo Baccigalupi, Enrico Cappellaro, Bepi
Tormen and Massimo Turatto for useful discussions.  We are grateful to
the referee, Peter Nugent, for useful suggestions in clarifying the
presentation of the results.}

\end{document}